\documentclass[aps,prl,twocolumn,superscriptaddress,longbibliography]{revtex4-1}
\usepackage[usenames, dvipsnames]{color}
\usepackage{graphicx}
\usepackage{color}
\usepackage{epstopdf}
\usepackage{epsfig}
\usepackage{bm}
\usepackage{xcolor}
\usepackage{amsmath}
\usepackage{amsfonts}
\usepackage{float}
\usepackage{subfigure}
\usepackage{verbatim}
\usepackage[version=3]{mhchem}
\usepackage[ansinew]{inputenc}
\usepackage{epstopdf}
\usepackage{scalerel}
\usepackage{multirow}
\usepackage{booktabs}
\usepackage{hyperref}
\usepackage[para,flushleft]{threeparttable}

\setcitestyle{super}

\newcommand{\degree}{^{\circ} }

\newcommand{\kT}{\ensuremath{k_{\rm B}T}}

\newcommand{\dgf}{\ensuremath{\Delta G_f}}
\newcommand{\dhf}{\ensuremath{\Delta H_f}}
\newcommand{\ds}{\ensuremath{\Delta S}}

\newcommand{\camno}{CaMnO$_3$}
\newcommand{\mgten}{Ca$_{0.9}$Mg$_{0.1}$MnO$_3$}

\newcommand{\srten}{Ca$_{0.9}$Sr$_{0.1}$MnO$_3$}

\newcommand{\feten}{CaMn$_{0.9}$Fe$_{0.1}$O$_3$}

\newcommand{\alten}{CaMn$_{0.9}$Al$_{0.1}$O$_3$}

\newcommand{\mnp}{Mn$^{4+}$}
\newcommand{\delm}{\ensuremath{\delta_{m}}}
\newcommand{\delb}{\ensuremath{\delta_{b}}}
\newcommand{\dev}{\ensuremath{\Delta E_{\rm vac} (\delta)}}
\newcommand{\devs}{\ensuremath{\Delta E_{\rm vac} (\bar{\delta})}}

\newcommand{\dhvs}{\ensuremath{\Delta H_{\rm vac} (\bar{\delta})}}

\begin{document}


\title{Oxygen vacancies beyond the dilute limit in doped CaMnO$_3$ perovskites and implications for screening materials in thermochemical applications}



\author{Harender S. Dhattarwal}
\email[]{harender.dhattarwal@rutgers.edu}
\affiliation{Department of Chemistry and Chemical Biology, Rutgers University, Piscataway, NJ, 08854, USA}
\author{Colin M Hylton-Farrington}
\affiliation{Department of Chemistry and Chemical Biology, Rutgers University, Piscataway, NJ, 08854, USA}
\author{Ian G. McKendry}
\altaffiliation[Present address: ]{Nascent Materials, Inc., Newark, NJ, 07102, USA}
\affiliation{HiT Nano, Bordentown, NJ, 08505, USA}
\author{Christopher Abram}
\email[Previously at. Current contact: ]{christopher.abram@hotmail.com}
\affiliation{HiT Nano, Bordentown, NJ, 08505, USA}
\affiliation{Department of Mechanical and Aerospace Engineering, Princeton University, Princeton, NJ, 08540, USA}
\author{Richard C. Remsing}
\email[]{rick.remsing@rutgers.edu}
\affiliation{Department of Chemistry and Chemical Biology, Rutgers University, Piscataway, NJ, 08854, USA}




\begin{abstract}
%
Thermochemical energy storage (TCES) in oxide perovskites relies on reversible oxygen vacancy formation, 
and computational high-throughput screening of candidate materials has predominantly used the single oxygen vacancy formation energy (OVFE) as the key descriptor. 
Here, we demonstrate that this descriptor is 
insufficient for screening cubic \camno~perovskites, 
because the stoichiometric compound is not the minimum energy reference state, 
and oxygen vacancies are inherently present at operating temperatures, which leads to negative single OVFE.
Materials with negative single OVFEs are routinely excluded from screening datasets as unsuitable, yet this exclusion reflects a mischoice of reference state rather than a genuine materials limitation, 
and risks discarding promising TCES candidates.
We address this by systematically computing OVFEs as a function of vacancy concentration using ab initio density functional theory, 
establishing the equilibrium vacancy concentration as the physically correct reference point. 
OVFE curves referenced to this minimum align closely with experimentally measured reduction enthalpies, providing a framework directly comparable to experiments. 
We further show that $A$-site doping (Mg, Sr) and $B$-site doping (Fe, Al) modify the vacancy formation landscape through distinct mechanisms.
$A$-site dopants act primarily through strain relaxation and symmetry breaking, while $B$-site dopants reshape the local redox environment and introduce strong configurational dependence. 
Finally, we develop a thermodynamic model incorporating configurational entropy that accurately predicts equilibrium oxygen stoichiometry as a function of temperature and oxygen partial pressure, 
and reveals that selective reduction of \mnp~versus $B$-site dopant ions can tune the onset temperature for vacancy formation. 
These results establish a physically rigorous screening framework for perovskite TCES materials and provide practical guidance for extending high-throughput workflows beyond the single-vacancy paradigm.

\end{abstract}


\maketitle

\raggedbottom

\section{Introduction}
\label{intro}

Increased emission of greenhouse gases is the leading cause of drastic climate change around the globe \cite{ipcc2023,creutzig2019mutual,rogelj2018mitigation}. 
This demands switching our energy and material requirements from fossil fuels to sustainable energy sources \cite{sims2004renewable,elshkaki2023implications}.
However, renewable energy sources like wind and solar face challenges in consistent power generation due to their dependence on variable factors including weather conditions, geographical location, and time of day \cite{ueckerdt2015analyzing,heptonstall2021systematic}.
Typically, excess electricity from renewable sources needs to be stored to provide energy when it is needed \cite{denholm2010role,lund2015review}, necessitating the develop of varied storage technologies, including electrochemical, geological, and thermal storage~\cite{dunn2011electrical,pardo2014review}.
Thermochemical energy storage (TCES) has emerged as a compelling strategy for addressing the intermittency of renewable energy sources, 
which stores thermal energy via reversible chemical reactions in addition to energy stored as sensible heat \cite{COTGORES2012,TCES_review2019,pardo2014review}. 
Thermochemical storage media have been proposed in concentrating solar power plants to increase the volumetric storage density \cite{TCES_Review_2016,TCES_review2019}. 
Reversible thermochemical reactions can also be used to exchange oxygen to drive two-step solar cycles for the production of synthetic fuels, for instance, for water splitting and CO$_2$ reduction \cite{bulfin2017applications,TWS_2015,Miller2013,Scheffe2014}. 
Thermochemical storage also enables long-duration storage, where reduced materials may be stored cold indefinitely and energy is recovered by heating the material, potentially enabling low-cost seasonal storage \cite{bulfin2015thermodynamics}.
Operating temperatures compatible with high temperature processes and concentrated solar power plants make metal oxides or mixed metal oxides such as CeO$_2$, Mn$_2$O$_3$, and Mn$_2$O$_3$ + Fe$_2$O$_3$ suitable candidates for TCES\cite{BLOCK2016,MOxides_2022,takacs2017splitting,MnO2014}.
Among candidate TCES materials, oxide perovskites ($AB$O$_3$) have attracted significant attention due to their continuous oxygen release, reversible and fast redox kinetics, and stability at higher temperatures~\cite{mcdaniel2013sr,bakken2005nonstoichiometry,scheffe2013lanthanum,nalbandian2009la,BAYON2020,Carter2021,Haile2021}. 
When oxide perovskites are heated to high temperatures, they undergo reduction leading to the formation of oxygen vacancies. 
Equilibrium is determined by the temperature and the oxygen partial pressure. 
The reduced perovskites store energy in chemical bonds, in addition to the sensible heat. 
This increases the energy density of storage, and allows the material to store energy for an indefinite period at ambient temperatures, if cooled in an oxygen deficient environment~\cite{bulfin2017redox}. 
Computational screening has played an increasingly important role in guiding the discovery and optimization of perovskite oxides for TCES and related thermochemical applications \cite{emery2017high,baldassarri2023oxygen,cai2023accelerated,emery2016high,Carter2021,Fanxing2022}.
Most high-throughput and descriptor-based studies estimate redox activity using the oxygen vacancy formation energy (OVFE), typically computed for the creation of a single, isolated oxygen vacancy in an otherwise stoichiometric lattice\cite{TCES_Review_2016,cai2023accelerated}.
The following redox reaction defines single oxygen vacancy formation in a bulk perovskite supercell containing \textit{n} unit cells, 
\begin{equation}
    nAB{\rm O_3} \longrightarrow nAB{\rm O}_{3-1/n} + \frac{1}{2} {\rm O}_2 {\rm (g)}.
\end{equation}
The OVFE of this redox reaction can be computed as \ensuremath{E_{AB{\rm O}_{3-1/n}}-E_{AB{\rm O}_3} + E_{{\rm O}_2}}/2.  
Here \ensuremath{E_{AB{\rm O}_{3-1/n}}} and \ensuremath{E_{AB{\rm O}_3}} are the total energies of the reduced and oxidized perovskite supercells, respectively, and \ensuremath{E_{{\rm O}_2}} is the energy of the oxygen molecule in the gas phase.
This dilute defect approximation has been adopted widely because of its computational efficiency and apparent correlation with experimentally measured reduction enthalpies in some systems\cite{mastronardo2020constantOVFE,vieten2017redox,bulfin2017redox}. 
As a result, materials exhibiting negative or very small single vacancy formation energies are often excluded from screening datasets under the assumption that such materials are unstable or unsuitable for reversible thermochemical cycling\cite{cai2023accelerated}.
%

\begin{figure*}[htb]
\begin{center}
\includegraphics[width=0.75\textwidth]{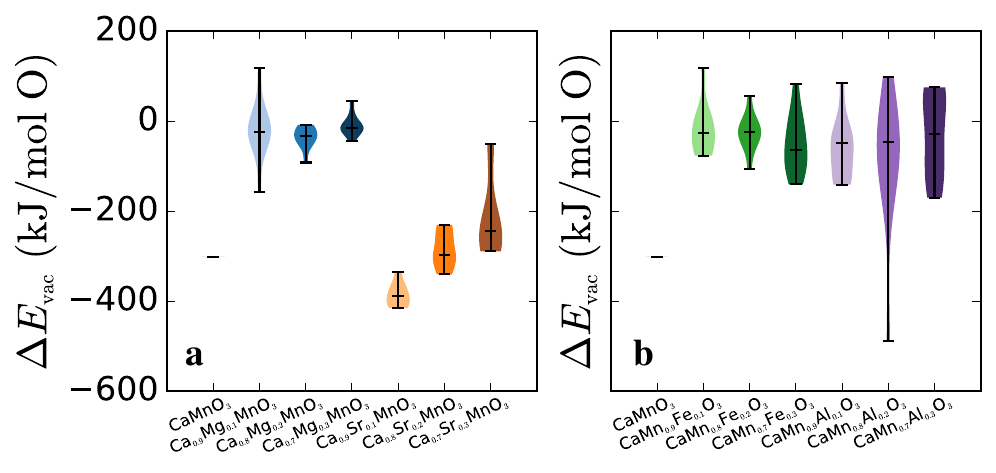}
\caption{\label{fig:ovfe} Violin plot for oxygen vacancy formation energy in \camno~when doped with different concentrations of (a) Mg and Sr at $A$-site, and (b) Fe and Al at $B$-site. Wider plot indicates higher probability of OVFE.}
\end{center}
\end{figure*}

Calcium manganite (\camno) provides a particularly instructive case study. 
While \camno~adopts an orthorhombic structure at room temperature, it undergoes a sequence of phase transitions to tetragonal and 
ultimately cubic symmetry at elevated temperatures (above 913 $\degree$C) relevant for TCES operation\cite{taguchi1989highcubic,CaMnFeO3_2019}. 
Experimental studies have shown that cubic \camno~readily accommodates oxygen vacancies and exhibits significant intrinsic oxygen deficiency under reducing conditions. 
Despite this, many computational studies either focus on the low temperature orthorhombic phase\cite{Huang2019ortho,Mishra2019}, where oxygen vacancy formation is low, or evaluate oxygen vacancy formation in the dilute limit, 
potentially overlooking the true thermodynamic driving forces governing oxygen release in the high temperature cubic phase.

In this work, we systematically investigate oxygen vacancy formation in cubic \camno~beyond the dilute defect approximation by explicitly computing vacancy formation energies as a function of oxygen vacancy concentration. 
Using first principles density functional theory (DFT), we show that the formation energy of a single oxygen vacancy in cubic \camno~is negative, indicating that the stoichiometric compound is not the true energetic reference state at high temperature.
Instead, the most stable configurations inherently contain a finite concentration of oxygen vacancies. 
This observation demonstrates that single vacancy formation energies alone are insufficient, and in some cases qualitatively misleading, as screening metrics for thermochemical applications.

We further examine how doping at the $A$-site and $B$-site modifies the vacancy formation landscape, focusing on representative dopants commonly explored in experimental TCES studies\cite{DopingCM2021,bulfin2017redox}. 
By computing vacancy formation energies relative to the equilibrium non-stoichiometric reference state, we establish a physically meaningful framework for comparing redox energetics across compositions. 
This approach enables direct comparison with experimentally measured reduction enthalpies, which inherently probe changes in oxygen content relative to the prepared material rather than the hypothetical stoichiometric limit.

Finally, we introduce a thermodynamic model that combines first principles energetics with configurational entropy to predict equilibrium oxygen stoichiometry as a function of temperature, oxygen partial pressure, and dopant concentration.
This model captures experimentally observed trends in redox onset temperature and oxygen release behavior, while remaining computationally tractable. 
More broadly, our results highlight the importance of incorporating vacancy concentration dependent energetics into computational screening workflows and suggest practical strategies for extending high-throughput studies beyond the single vacancy paradigm.

\section{Methods}
First principles DFT simulations were executed using the Vienna ab initio Simulation package software (VASP)\cite{kresse1996VASP1}. 
We used $3\times3\times3$ perovskite supercells, which provide sufficient separation between periodic vacancy images and allow concentrations of oxygen vacancies, $\delta$, ranging from approximately 0.037 (one vacancy) to 0.44 (12 vacancies out of 27 available oxygen sites per formula unit). 
At these supercell sizes, finite-size electrostatic corrections for neutral supercells are not expected to contribute significantly to the OVFE trends.
We utilized the frozen core all-electron projector augmented wave model and the Perdew-Burke-Ernzerhof functional\cite{kresse1999VASP2,perdew1996generalized}.
The energy cut off was set to 520~eV, and we ensured that the force and energy converged with the criteria of 0.01~eV~\AA$^{-1}$ and 10$^{-5}$~eV, respectively. 
For optimization simulations, Gaussian smearing was employed with a width of 0.05~eV. We used a $3\times3\times3$ $k$-point grid for the perovskite supercells. 
Additionally, the DFT + U method was utilized for the $d$ orbitals of Fe and Mn with U values of 4 and 3.8~eV, respectively\cite{anisimov1991bandU}. 
The parameters for Hubbard U corrections optimized for VASP provide good estimates of the electronic structure of perovskite materials \cite{Hong_U2012,ParamU2018}.

The high thermal conductivity makes vacancy formation possible throughout the bulk. Therefore, bulk OVFE calculations should provide good estimates on the TCES performance of these materials. 
Vacancies formed in bulk materials are quite stable and usually do not diffuse to the surface\cite{BulkVacancyML_2023}. Therefore, we kept the system periodic in three dimensions and studied bulk oxygen vacancies. 
We used a quasi-random Monte Carlo method to introduce dopants and vacancies in the lattice. The site of oxygen vacancy does not affect OVFE significantly in unstrained perovskites\cite{PRB2013}. However, we still ran 10 separate calculations to estimate the oxygen vacancy formation energy in the system.

Undoped and doped \camno~samples were prepared using a solid state method from metal oxide and carbonate precursor powders purchased from generic scientific chemical suppliers. Precursor powders were mixed together using a pestle and mortar and pressed into half-inch diameter disc-shaped pellets using a die press.  Pellets were fired on alumina plates inside a box furnace at 1350$\degree$ C and ground again for analysis. 

XRD was used to determine the crystal phase of undoped CaMnO$_3$ as orthorhombic. A modified van't Hoff method following that of Babinec et al. \cite{BABINIEC2015} was used to measure the thermochemical reaction enthalpy of the samples. Samples were loaded into a  TG-DSC/DTA (Netzsch STA 449 F5 Jupiter) and subjected to long heating and cooling ramps in various gas atmospheres (0.5\%, 5\%, 21\% and 75\% O$_2$). Mass losses during cycling correspond to oxygen deficiency  $\delta$. Plots of  $\delta$ (assuming any pre-existing oxygen deficiency is baselined to $\delta = 0$) against temperature could then be extracted for various oxygen partial pressures, from which enthalpy data can be determined and integrated to quantify thermochemical storage enthalpies.

\begin{figure*}[htb]
\begin{center}
\includegraphics[width=0.75\textwidth]{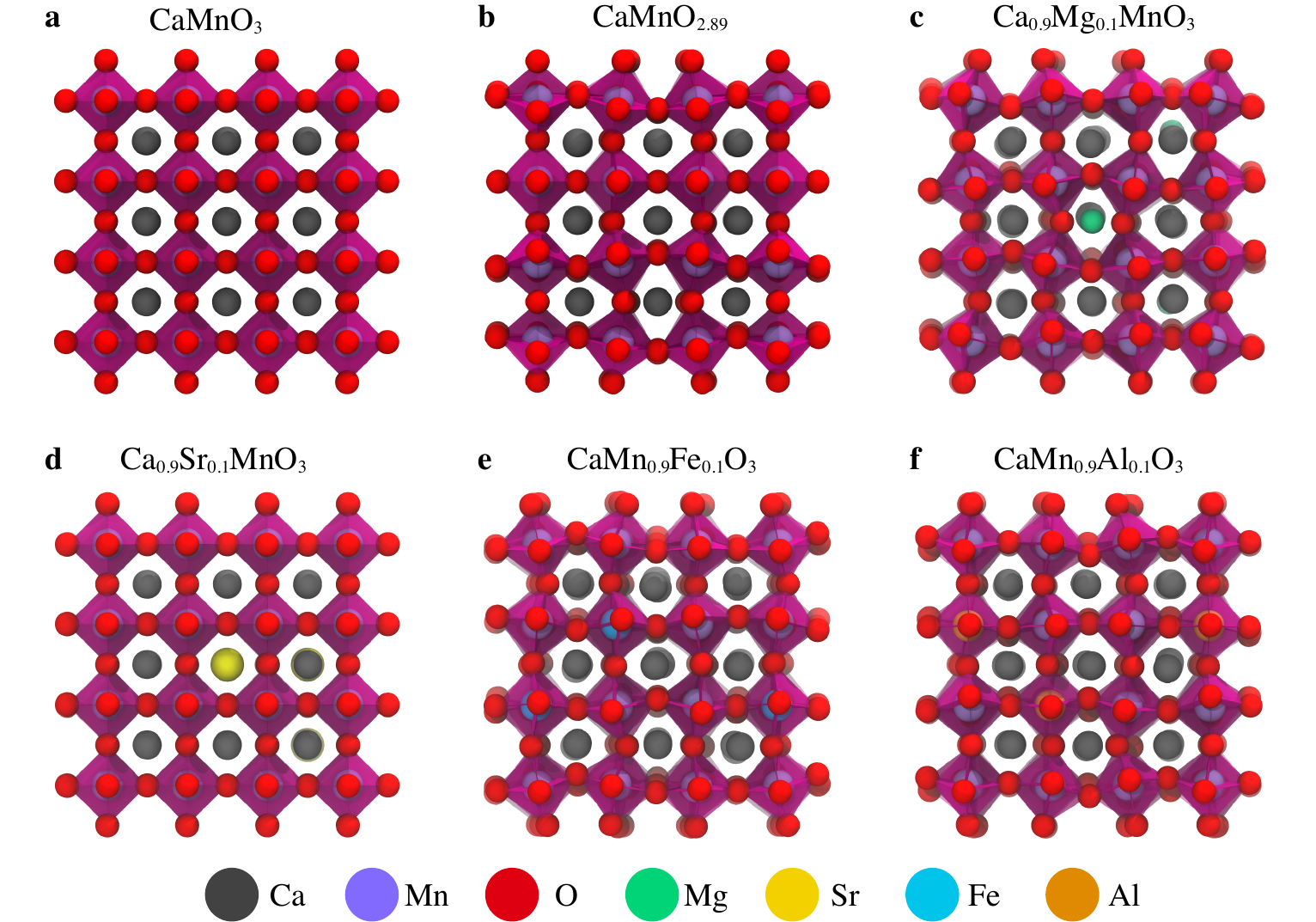}
\caption{\label{fig:snaps} Snapshots showing the equilibrated structure of (a) \camno~without any oxygen vacancy, (b) \camno~with single oxygen vacancy, (c) \mgten, (d) \srten, (e) \feten, and (f) \alten~without any oxygen vacancy.}
\end{center}
\end{figure*}

\begin{figure*}[htb]
\begin{center}
\includegraphics[width=0.75\textwidth]{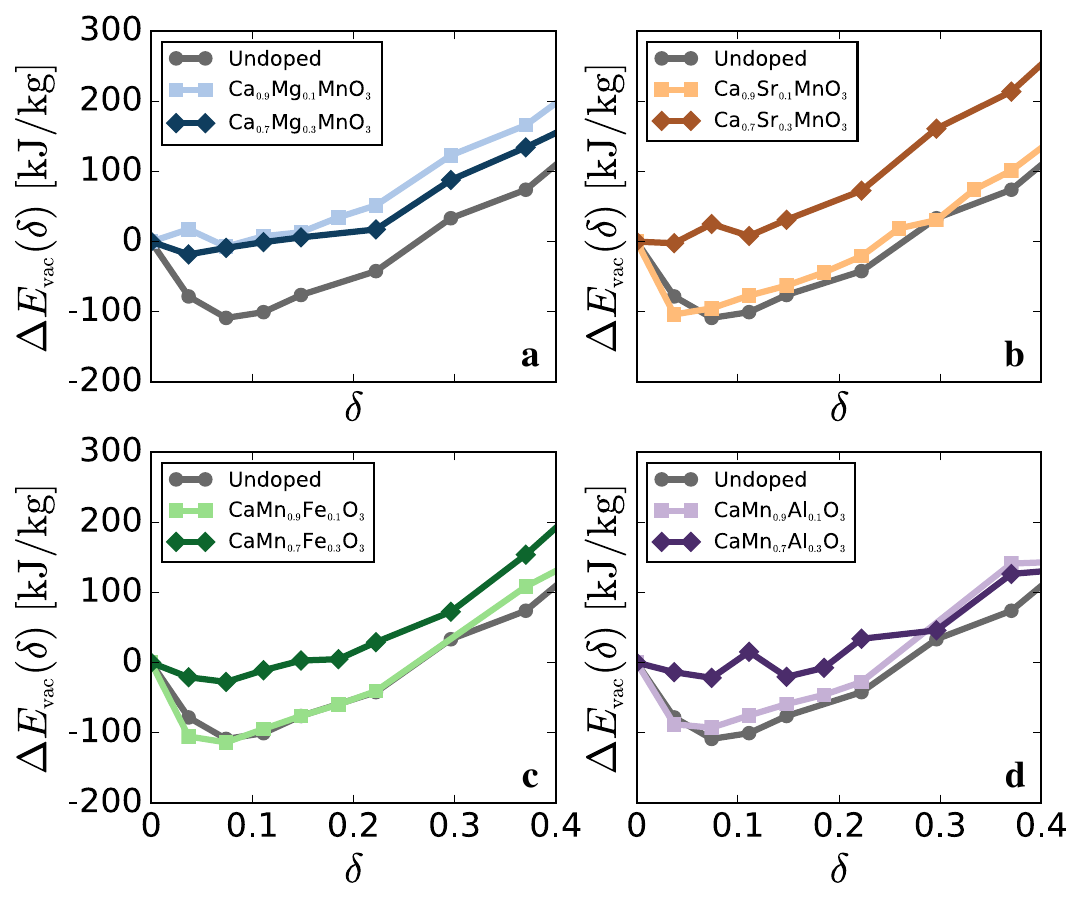}
\caption{\label{fig:fullovfe} Oxygen vacancy formation energy in \camno~when doped with different concentrations of (a) Mg and (b) Sr at $A$-site, and (c) Fe and (d) Al at $B$-site.}
\end{center}
\end{figure*}

\begin{figure*}[!htb]
\begin{center}
\includegraphics[width=0.75\textwidth]{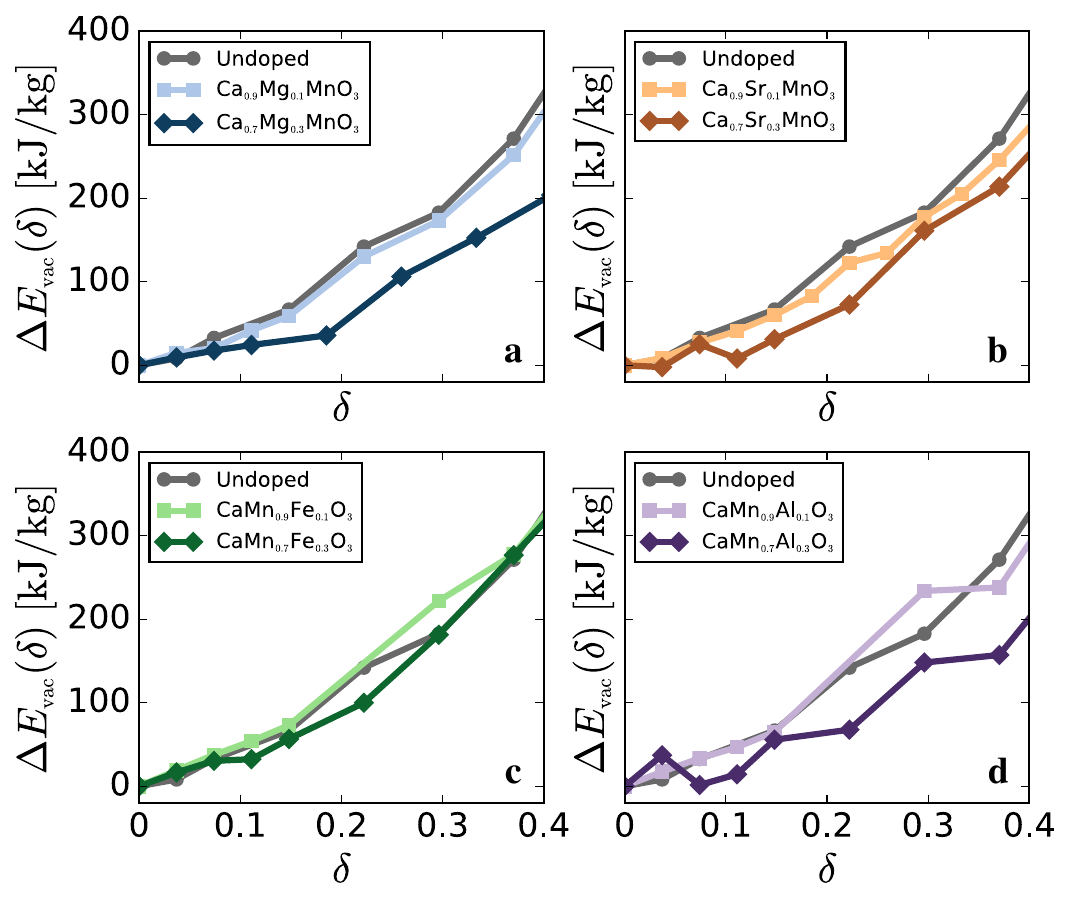}
\caption{\label{fig:shiftedovfe} Oxygen vacancy formation energy in \camno~when doped with different concentrations of (a) Mg and (b) Sr at $A$-site, and (c) Fe and (d) Al at $B$-site, referenced with respect to the equilibrium vacancy concentration ($\bar\delta = \delta - \delta_{\rm eq}$).}
\end{center}
\end{figure*}

\section{Results and Discussion}\label{sec:rnd}

\subsection{Single oxygen vacancy formation energy}
\camno~adopts an orthorhombic structure at lower temperatures and in oxygen rich environments\cite{CubicCAMNO3_2022}.
In the orthorhombic phase, \camno~requires high energy to create oxygen vacancies\cite{OrthoCMO2020,PRB2013}. 
However, our goal here is to estimate the effect of dopants at operational temperatures of thermochemical energy storage.
Hence, we have studied cubic \camno, which is the phase present at the high temperatures of interest in this study.
In Fig. \ref{fig:ovfe}, we show single OVFEs for cubic \camno~perovskites with different dopants at the $A$ (Ca$_xA_{1-x}$MnO$_3$) and $B$ (CaMn$_yB_{1-y}$O$_3$) sites. 
Our calculations show that the OVFE in pristine \camno~is negative, indicating that oxygen removal from cubic \camno~is energetically favorable. 
The energetic preference for vacancy formation in the cubic phase is due to the high symmetry of the cubic lattice, (Fig. \ref{fig:snaps}a and \ref{fig:snaps}b), which lacks the rigid, cooperative octahedral tilt patterns that stabilize the low-temperature phase\cite{PRB2013,marthinsen2016coupling}.
Unlike the orthorhombic form, where removing an oxygen atom disrupts optimized bond networks and incurs a high elastic penalty, the cubic framework can more readily absorb local relaxations and accommodate the electronic redistribution associated with reducing Mn$^{4+}$ to Mn$^{3+}$. 
The cubic phase becomes thermodynamically stabilized by the lower enthalpy of vacancy formation and the increased configurational entropy.
This corroborates the experimental findings that cubic \camno~at higher temperatures inherently contains oxygen vacancies\cite{CaMnFeO3_2019}. 
The reduction enthalpy of \camno~perovskites decreases after transition to the tetragonal or cubic phase at higher temperatures.

Doping with different elements at the $A$-site can influence the strain in the lattice, which can affect the OVFE.
We introduced a smaller atom (Mg) and a larger atom (Sr) at the $A$-site to study the effects of dopant size on OVFE (Fig. \ref{fig:ovfe}a).
The vertical spread of the plot corresponds to OVFE values for different configurations (random distribution of dopants over $A$-sites) of the same dopant concentration.
Doping the $A$-site with 10\% Mg significantly increases the OVFE. 
Once \camno~is doped with Mg, the reference state is no longer a high symmetry, overconstrained lattice. 
Even before introducing oxygen vacancies, dopants lower the symmetry and activate octahedral tilts and local $A$-site off centering (see Fig. \ref{fig:snaps}c).
The lattice has already paid the symmetry breaking cost and relieved a substantial fraction of elastic strain. 
As a result, oxygen removal from this pre-relaxed structure provides a smaller additional energy benefit, leading to an increased OVFE.
The horizontal spread of the plot indicates the probability of OVFE values for different configurations, all indicating higher OVFE than undoped \camno.
The OVFE does not increase significantly upon subsequent doping with Mg. 

In contrast to smaller atom (Mg) doping, substitution with larger atom (Sr) at the $A$-site does not significantly alter the OVFE relative to pristine \camno~(Fig. \ref{fig:ovfe}a).
Unlike Mg substitution, Sr doping does not introduce pronounced octahedral tilts or local distortions in the stoichiometric structure (Fig. \ref{fig:snaps}d). 
The reference state remains a high-symmetry, relatively unstrained framework similar to pristine cubic \camno~and the structural and electronic response upon oxygen removal closely resembles that of the undoped system. 
As a result, the energetic driving force for vacancy formation remains comparable to that of pristine \camno, and only a modest increase in OVFE is observed with increasing Sr content.
Alternatively to relaxing lattice strain and lowering symmetry through $A$-site substitution, cation substitution at the $B$-site modifies the redox chemistry of oxide perovskites, which can have a pronounced impact on the OVFE\cite{Steinfeld2015}. 
To examine these effects, we doped \camno~at the $B$-site with Fe, a redox-active transition metal, and Al, a redox-inactive post-transition metal (Fig. \ref{fig:ovfe}b).
In contrast to $A$-site doping, $B$-site substitution leads to a substantially broader distribution of OVFE values. 
This enhanced spread reflects a strong dependence of OVFE on the relative positions of the dopant and the oxygen vacancy, as oxygen removal now occurs from chemically distinct local environments involving Mn-O-Mn, Mn-O-$B$, or $B$-O-$B$ linkages. 
As a result, the OVFE varies significantly across different configurations at the same dopant concentration.
Despite this broad distribution, the median OVFE for most $B$-site doped systems remains higher than that of undoped \camno, indicating that oxygen vacancy formation is comparatively less favorable once the MnO$_6$ network is chemically perturbed. 

Overall, these results highlight a qualitative difference between $A$- and $B$-site doping. While $A$-site dopants primarily influence OVFE through global strain relaxation and symmetry breaking,
leading to relatively uniform shifts in OVFE,
$B$-site dopants reshape the local bonding and redox landscape of the perovskite. 
This leads to strong configurational dependence and a wide range of OVFE values, underscoring the dominant role of $B$-site chemistry in controlling oxygen vacancy formation in \camno\ perovskites.

Despite significant changes, the overall magnitude of the vacancy formation energy is still small (or negative) for most configurations of doped systems.
We will later show that the computational estimate of a single OVFE is not necessarily a good indicator of real world performance of these materials for TCES applications. 
Usually, materials with negative OVFE are neglected in high-throughput screening for thermochemical applications\cite{cai2023accelerated}. In the world of machine learning, careful data labeling is extremely important, and excluding such materials can overlook potential TCES candidates.

\subsection{Beyond single oxygen vacancy formation}

The operating temperatures of the thermal reduction process of perovskites are usually higher than 1000 K.
At such high temperatures, these materials have a high concentration of oxygen vacancies. Therefore, it is important that we investigate the creation of many oxygen vacancies. 
Here, we analyzed OVFEs at varying concentrations of oxygen vacancies in \camno~perovskites as
\begin{equation}
    \Delta E_{\rm vac} (\delta)=E_{{\rm ABO}_{3-\delta}}-E_{{\rm ABO}_3} + \frac{\delta}{2} E_{{\rm O}_2},
\end{equation}
where $\delta$ is the concentration of oxygen vacancies in the perovskite (Fig. \ref{fig:fullovfe}).
For the undoped \camno~system, the initial negative \dev~at low $\delta$~arises from the same mechanism discussed earlier for the single vacancy case. 
The ideal cubic phase is mechanically overconstrained, and the first few oxygen vacancies act as symmetry-breaking defects that unlock large octahedral relaxations. These early vacancies therefore provide a net energetic gain. 
These negative \dev~values suggests that the most stable configurations of cubic \camno~inherently contain vacant oxygen sites.

As $\delta$ increases, however, the lattice progressively loses its cubic character and octahedral tilts become established, and the elastic degrees of freedom that initially stabilized vacancy formation are largely exhausted. 
Beyond this point, additional oxygen removal primarily breaks already relaxed Mn-O bonds and introduces electrostatic and chemical repulsion between nearby vacancies, leading to a monotonic increase in \dev~with increasing $\delta$.
This behavior is specific to the cubic reference state. In orthorhombic \camno, where octahedral tilts are already present in the stoichiometric structure, the first oxygen vacancy does not unlock comparable relaxation pathways, and the \dev~is positive from the outset\cite{PRB2013,Huang2019ortho}.

For the doped systems, the situation is fundamentally different. In all doped cases, the stoichiometric reference structures are already symmetry broken and locally relaxed due to the presence of dopants. 
Consequently, the first few oxygen vacancies do not act as uniquely destabilizing or stabilizing perturbations, 
and the energetic response to increasing $\delta$ depends sensitively on the specific local arrangement of dopants and vacancies.

As a result, no universal trend in \dev~is observed across doped systems. 
Instead, the \dev~reflects a competition between local strain relief, dopant-vacancy interactions, and electronic compensation mechanisms that vary from one configuration to another. 
Since the present calculations were performed for a single representative dopant-vacancy configuration at each concentration, the resulting trends should be interpreted as configuration specific rather than statistically averaged behavior.
This underscores the importance of configurational sampling when assessing defect thermodynamics in doped perovskite oxides at high oxygen non-stoichiometry.

\begin{figure}[htb]
\begin{center}
\includegraphics[width=3.0in]{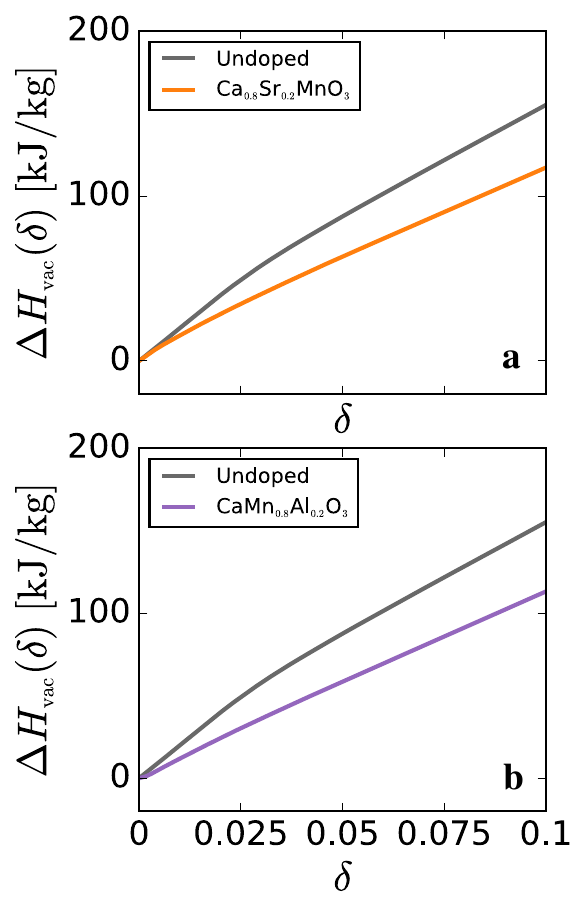}
\caption{\label{fig:expovfe} Experimentally measured oxygen vacancy formation energy in \camno~when doped with (a) Sr at $A$-site and (b) Al at $B$-site.}
\end{center}
\end{figure}

Experiments estimate $\delta$ based on the loss in mass from oxygen removal, which does not account for any pre-existing vacancies.
Hence, the experimentally estimated $\delta$ represents change in oxygen stoichiometry rather than the absolute vacancy concentration. 
The minimum in \dev~would be the initial state of the materials in the experimental setup. 
Therefore, we have adjusted \dev~curves to reference the minima in the energy trends, as depicted in Figure \ref{fig:shiftedovfe},
\begin{equation}
    \Delta E_{\rm vac} (\bar{\delta})=E_{{\rm ABO}_{3-\delta}}-E_{{\rm ABO}_{3-\delta_{\rm eq}}} + \frac{\bar{\delta}}{2} E_{{\rm O}_2},
\end{equation}
$\bar{\delta}$ is $\delta$ - $\delta_{\rm eq}$, where $\delta_{\rm eq}$ is the value of delta at the minimum (the equilibrium vacancy concentration).

We then compared our computational predictions with the enthalpy of oxygen vacancy formation (\dhvs) obtained from experiments (shown in Figure \ref{fig:expovfe}). 
Experimental \devs~decreased slightly as we introduced dopants at $A$- and $B$-sites. These results are consistent with previous experimental reporting of decreased enthalpy at a given $\bar{\delta}$ value upon doping with Fe at $B$-site.\cite{CaMnFeO3_2019}
The \dev~calculations suggested that the OVFEs were higher in \camno~doped by Sr and Al and $A$- and $B$-sites. 
However, the shift in the reference point revealed that the introduction of dopants led to a slight decrease in the \devs. Further decrease was observed when we increased the dopant concentration to 30\%. Interestingly, this trend aligns more closely with experimental findings, which also suggest a decrease in enthalpy at higher dopant concentrations. 
The developed computational framework for multiple oxygen vacancies 
supports the general conclusion that the role of doping does not necessarily lead to gains in enthalpy due to appreciable increases in OVFE for the same oxygen concentration and temperature. Doping improves storage performance by stabilizing the compound to allow a deeper reversible reduction at higher temperatures.
The adjustment of \dev~with respect to minima not only improves the accuracy of our predictions but also provides a more realistic representation of the behavior of oxygen vacancies in \camno~perovskites.

In our calculations, \dev~curves are somewhat nonlinear with $\delta$. 
The slopes of the energy curves are steeper at higher oxygen vacancy concentrations, indicating that the successive oxygen removal from the perovskite lattice requires more energy.
A similar trend is also observed for SrFeO$_3$, where the OVFE is higher at higher $\delta$ values\cite{MultipleVacancy2_2021}.
This could be because at higher oxygen vacancy concentrations, the cations near the oxygen vacancies are already reduced, leading to a higher energy requirement for further reduction.
Although, if the dilute approximation for vacancies is considered, \textit{i.e.}, vacancies are far from each other, subsequent oxygen vacancies can be created with same amount of energy\cite{MultipleVacancy1_2020}. 

\subsection{Thermodynamic model for the oxygen deficiency}

Although our calculations provide a qualitative estimate of the effect of dopants on vacancy formation, the calculated \devs~are significantly lower than experimental values. This discrepancy between our computational results and experimental data underscores the complexity of the system and the challenges involved in accurately predicting its behavior. 
One reason could be that the simulations neglect thermal contributions. 
Below, we describe a thermodynamic model to obtain the molar oxygen content (3-$\delta$) as a function of temperature. 
We aim to use this analysis to predict the onset temperature (when the oxygen vacancies start to form at a significant rate) directly from the reaction enthalpies.

\begin{figure}[htb]
\begin{center}
\includegraphics[width=0.45\textwidth]{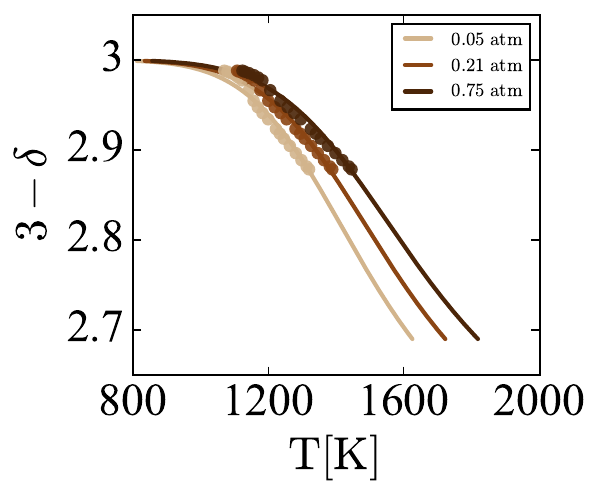}
\caption{\label{fig:deltaT} Model oxygen stoichiometry of \camno~at different oxygen partial pressures as a function of temperature. Here, lines represent our predictions, and solid circles represent experimental values.}
\end{center}
\end{figure}

\begin{figure*}[htb]
\begin{center}
\includegraphics[width=0.75\textwidth]{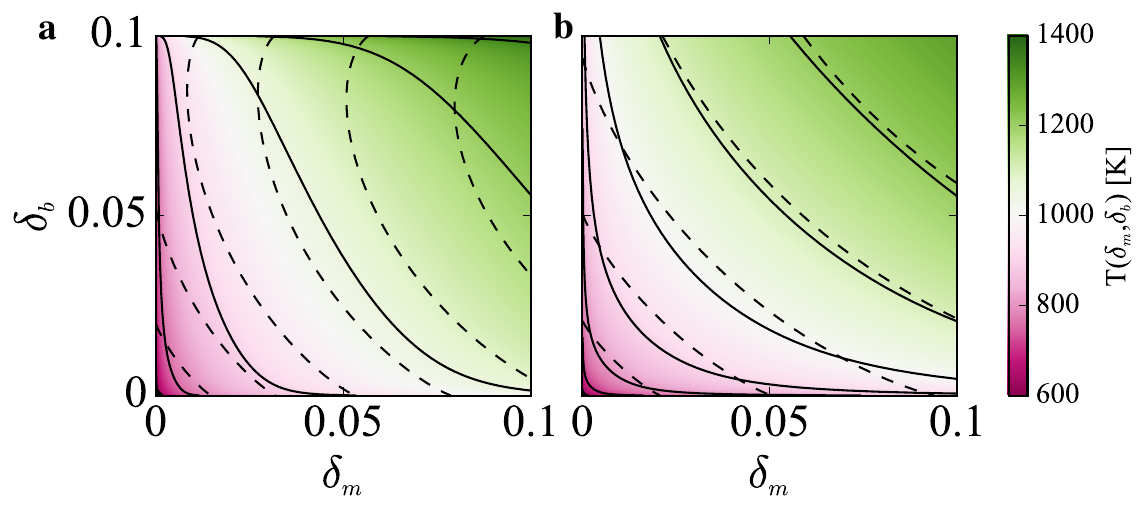}
\caption{\label{fig:Tdeltadelta} The temperature required to form oxygen vacancies in \camno~with 
(a) 20\% and (b) 50\% of the $B$-sites doped with Al as a function of the relative distribution of reduction sites over \mnp~and dopant ions.
The dashed lines show the trends in entropy associated with the corresponding vacancy concentration.
}
\end{center}
\end{figure*}

Usually, \mnp~at the $B$-site are reduced during oxygen vacancy formation\cite{OrthoCMO2020}. 
The chemical reaction for oxygen vacancy ($V_{{\rm O}}$) formation in undoped-\camno~is 
\begin{equation}
\label{eqn:undoped}
     4{\rm Mn}^{4+} + 2{\rm O}^{2-} \longrightarrow 4{\rm Mn}^{3+} + 2V_{{\rm O}} + {\rm O}_{2} (g).
\end{equation}
The equilibrium constant, $K$, for this reaction is
\begin{equation}
\label{eqn:eqbm1}
    K = \frac{[{\rm Mn}^{3+}]^{4} [V_{{\rm O}}]^{2} p_{{\rm O}_{2}}}{[{\rm Mn}^{4+}]^{4}[{\rm O}^{2-}]^{2}} = {\rm exp}\left( -\frac{\dgf}{\kT} \right).
\end{equation}
At equilibrium, {$[{\rm Mn}^{3+}] = 2\delta$}, 
$[V_{{\rm O}}] = \delta$, 
{$[{\rm Mn}^{4+}] = 1-2\delta$},  
and $[{\rm O}^{2-}] = 3-\delta$. 
\dgf~(= \dhf~$- T$\ds) is the Gibbs free energy of reaction \ref{eqn:undoped}. 
The above equation can be rearranged to predict oxygen stoichiometry as a function of temperature at any given oxygen partial pressure\cite{Bakken2002} (shown in Fig. \ref{fig:deltaT}). 
\dhf~was taken from our experimental measurements and the entropy was obtained from a fit at $p_{{\rm O}_2}=0.21$~atm. 
Then we predicted the oxygen stoichiometry at higher and lower oxygen partial pressures. Our calculation predicted that at a given temperature, it is easier to form oxygen vacancies at lower oxygen partial pressures. 
Although this trend is intuitive and is in good agreement with the experiments, our calculations also predicted the oxygen stoichiometry at $p_{{\rm O}_2}=0.05$~atm and  $p_{{\rm O}_2}=0.75$~atm with high accuracy.

\begin{figure*}[htb]
\begin{center}
\includegraphics[width=0.75\textwidth]{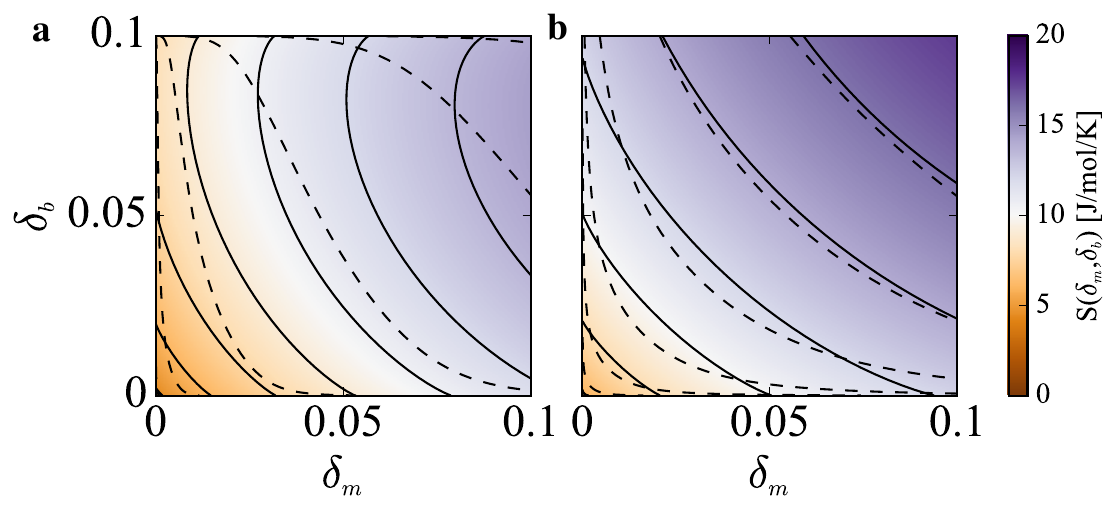}
\caption{\label{fig:Sdeltadelta} The entropy of oxygen vacancy formation in \camno~with 
(a) 20\% and (b) 50\% of the $B$-sites doped with Al as a function of the relative distribution of reduction sites over \mnp~and dopant ions.
The dashed lines show the trends in onset temperature associated with the corresponding vacancy concentration.
}
\end{center}
\end{figure*}

Further, we dope a fraction, $y$, of $B$-sites with a dopant, and assume that both sites have equal probability of getting reduced. The chemical reaction for the oxygen vacancy formation in doped-\camno~is 
\begin{align}
\label{eqn:reaction}
    4(1-y){\rm Mn}^{4+} + 4y{B}^{4+} + 2{\rm O}^{2-} &\longrightarrow 4(1-y){\rm Mn}^{3+}  \notag \\
    &+ 4y{B}^{3+} + 2V_{{\rm O}} + {\rm O}_{2} (g). 
\end{align}
The equilibrium constant for this reaction is
\begin{equation}
\label{eqn:eqbm}
    K = \frac{[{\rm Mn}^{3+}]^{4-4y} [{B}^{3+}]^{4y} [V_{{\rm O}}]^{2} p_{{\rm O}_{2}}}{[{\rm Mn}^{4+}]^{4-4y}[{B}^{4+}]^{4y}[{\rm O}^{2-}]^{2}} = {\rm exp}\left( -\frac{\dgf}{\kT} \right)
\end{equation}
At equilibrium, {$[{\rm Mn}^{3+}] = 2\delm$}, 
{$[{B}^{3+}] = 2\delb$}, 
$[V_{{\rm O}}] = \delta$, 
{$[{\rm Mn}^{4+}] = 1-y-2\delm$}, 
{$[{B}^{4+}] = y-2\delb$}, 
and $[{\rm O}^{2-}] = 3-\delta$. 
Similar to the undoped system, we can obtain the temperature required to form $\delta (= \delta_m + \delta_b)$ oxygen vacancies in the system as a function of dopant concentration 
and the relative distribution of reduction sites over \mnp~and dopant ions by rearranging the above equation. 
Previous estimates assume that equal number of \mnp~and dopant ions are reduced at the same time\cite{OrthoCMO2020}. 
However, in our analysis we considered distributing reduction sites between \mnp~and dopants. 
Fig.~\ref{fig:Tdeltadelta} shows the temperature, $T(\delm,\delb)$, required to reduce a concentration of \delm~\mnp~and \delb~dopant ions, resulting in a total oxygen vacancy concentration of $\delta$ in \camno~at 
$p_{{\rm O}_2}=0.21$~atm.
We focus on representative data with dopant concentrations of 20\% and 50\%. 
At a lower dopant concentration, the reduction of dopant ions is favored over that of \mnp, assuming that the reduction potential is the same for both ions. 
For example, $T(0.1, 0.0)\approx1100$~K while $T(0.0, 0.1)\approx900$~K, suggesting a preference for reducing the $B$-site dopants.
In addition, our predictions also suggest that reduction of a single ion type is favored over simultaneous reduction of both ion types, even at 50\% doping concentration. 
We also investigate the role of configurational entropy in oxygen vacancy formation. 
The configurational entropy of the reaction depends on the disorder of the oxygen vacancies in \camno, position of $B$-sites undergoing reduction, and the disorder in the dopant sites. 
\begin{align}
    \Delta S = &- R \bigg [ 
    (1 - 2\delta) {\rm ln} (1 - 2\delta) 
    + 2\delta {\rm ln}(2\delta) \notag \\
    & + (1 - y) {\rm ln} (1 - y) 
    + y {\rm ln}(y) \notag \\
    & + \delta{\rm ln}\left( \frac{\delta}{3} \right) 
      + (3 - \delta) \left( 1 - \frac{\delta}{3} \right) \bigg ]
\end{align}
Fig. \ref{fig:Sdeltadelta} depicts the configurational entropy, $S(\delm,\delb)$, corresponding to the reduction of \delm~and \delb~sites in the 20\% and 50\% $B$-site doped \camno~at $p_{O_2}=0.21$~atm. 
At 20\% dopant concentrations (Fig. \ref{fig:Sdeltadelta}(a)), the configurational entropy is significantly higher if \mnp~ions are selectively reduced, $S(0.1, 0.0) > S(0.0, 0.1)$. 
In contrast, At 50\% dopant concentration (Fig. \ref{fig:Sdeltadelta}(b)), the simultaneous reduction of \mnp~and $B$-site dopant ions is entropically favored, $S(0.05, 0.05)> S(0.1, 0.0) \approx S(0.0, 0.1)$. 
Our results suggest that selective doping of \camno~with ions of different reduction potentials can be used to tune the onset temperature. 
For example, introducing dopants with a reduction potential higher than \mnp~could lead to an increased concentration of oxygen vacancies at fixed temperature. 
In addition, our predictions regarding configurational entropy suggest that even higher vacancy concentrations will result from more reducible dopants. 
In principle, the thermodynamic model developed here is fully compatible with DFT-derived enthalpies. 
Substituting computed \devs~values corrected for vibrational contributions and magnetic ordering would render the framework entirely first-principles, enabling truly predictive screening without experimental thermochemical input. 
This represents a natural and tractable extension of the present work.

\section{Conclusions}
The oxygen vacancy concentration governs the performance of oxide perovskites in thermochemical energy storage.
Most computational studies focus on the single oxygen vacancy formation energy as an estimate to the ease of oxygen vacancy formation in these materials.
In this study, we have reported that single OVFEs are usually negative for undoped and doped cubic \camno, indicating that oxygen removal is energetically favorable.
Many previous computational studies rejected materials with negative OVFEs as unsuitable for TCES applications. 
However, the cubic \camno~inherently has vacant oxygen sites. The true estimate of OVFE should be referenced to the most stable configuration of \camno~with implicit oxygen vacancies.
Therefore, single OVFE may not be a good indicator of the TCES performance of \camno.

We addressed this by computing OVFE as a function of vacancy concentration ($\delta$). \dev~has a minima at low $\delta$ values corresponding to the most stable cubic \camno~with implicit oxygen vacancies.
Referencing \dev~to the most stable stoichiometry provided a more accurate description, and enabled us to compare our calculations directly with experiments. 
Our calculations showed that the energy of vacancy formation decreased slightly when doped with Sr at $A$-site and Al at $B$-site. These observations are supported by our experiments. 

We anticipate that performing calculations at finite temperatures will further improve the results, but require expensive simulations which are beyond the scope of the current study. 
We used a thermodynamic model to predict the effect of temperature and configurational entropy on the oxygen vacancy formation. 
Our model accurately predicted the change in oxygen stoichiometry with varying oxygen partial pressures. 
We also predicted from trends in configurational entropy and onset temperature that different concentrations of oxygen vacancies can be achieved by selective reduction of \mnp~or $B$-site dopant ions.
These insights can be helpful in fine tuning the perfomance of oxide perovskites for thermochemical applications, such as waste heat utilization, long-duration seasonal storage, and chemical production using concentrating solar-driven two-step
thermochemical cycles.

\section*{Conflicts of interest}
There are no conflicts to declare.

\section*{Acknowledgments}
This material is based upon work supported by the U.S. Department of Energy, Office of Science, SBIR program under Award Number DE-SC0020731

\section*{Data Availability}
The data generated in this study are presented in the main article and the Source Data file.


\bibliography{TCES}


\end{document}